\documentclass[aip,amsmath,amssymb,reprint,nobibnotes,noeprint]{revtex4-2}
\usepackage{graphicx,xcolor}
\usepackage{dcolumn}
\usepackage{bm}
\usepackage[utf8]{inputenc}
\usepackage[T1]{fontenc}
\usepackage{mathptmx}
\usepackage{physics}
\usepackage{hyperref}

\newcommand{\rms}{\rm\scriptscriptstyle}
\DeclareMathOperator*{\argmin}{argmin}

\begin{document}

\title[Driven transport of soft Brownian particles: Effective size method]{Driven transport of soft Brownian particles through pore-like structures: Effective size method}

\author{Alexander P.\ Antonov}
\email{alantonov@uos.de}
\affiliation{ 
Universit{\" a}t Osnabr{\" u}ck, 
Fachbereich Physik, 
Barbarastra{\ss}e 7, 
D-49076 Osnabr{\" u}ck, 
Germany}

\author{Artem Ryabov}
\email{rjabov.a@gmail.com}
\affiliation{Charles University, 
Faculty of Mathematics and Physics, 
Department of Macromolecular Physics, 
V Hole\v{s}ovi\v{c}k\'ach 2, 
CZ-18000 Praha 8, 
Czech Republic}

\author{Philipp Maass} 
\email{maass@uos.de}
\affiliation{ 
Universit{\" a}t Osnabr{\" u}ck, 
Fachbereich Physik, 
Barbarastra{\ss}e 7, 
D-49076 Osnabr{\" u}ck, 
Germany}


\begin{abstract}

Single-file transport in pore-like structures constitute an important topic for both theory and experiment. For hardcore interacting particles, a good understanding of the collective dynamics has been achieved recently.
Here we study how softness in the particle interaction affects the emergent transport behavior. To this end, we investigate the driven Brownian motion of particles in a periodic potential. The particles interact via a repulsive softcore potential with a shape corresponding to a smoothed rectangular barrier. This shape allows us to elucidate effects of mutual particle penetration and particle crossing in a controlled manner. We find that even weak deviations from the hardcore case can have a strong impact on the particle current. Despite this fact, the knowledge about the transport in a corresponding hardcore system is shown to be useful to describe and interpret our findings for the softcore case. This is achieved by assigning a thermodynamic effective size to the particles based on the equilibrium density functional of hard spheres.
\end{abstract}

\maketitle

\section{\label{sec:level1}Introduction}
Understanding transport through pore-like structures is important for many chemical and biophysical processes and applications.
Examples include pores of zeolites in catalysis,\cite{VanDeVoorde/Sels:2017} carbon nanotubes with
relevance to biotechnological and biomedical applications,\cite{Zeng/etal:2018} and particle motion in nanofluidic devices,\cite{Ma/etal:2015} 
which can be utilized for various needs such as water filtration and osmotic energy conversion. 

Frequently the confining environment for the particle transport has a periodic
structure that in theoretical modelings is  most easily accounted
for by an external periodic potential with the wavelength $\lambda$. This potential can be
of entropic origin, e.g., reflecting variations in a pore cross section,
\cite{Mon/Percus:2007, Burada/etal:2009, Mangeat/etal:2018} or/and of
energetic nature, e.g., when associated with binding sites inside a
membrane channel.\cite{Hilty/Winterhalter:2001, Kasianowicz/etal:2006, Goldt/Terentjev:2014} 
 
If the pore size is comparable to the particle diameter, the particle motion often has a single-file 
character.\cite{Hartmann:2005, Humphrey/etal:2001, Hille:2001, Hahn/etal:1996, Chmelik/etal:2018, Cheng/Bowers:2007, Dvoyashkin/etal:2014}
This means that the particles involved in the transport cannot pass each other and thus keep their 
ordering. The no-passing condition has severe implications on the dynamics. This was first realized in connection with 
anomalous subdiffusion of tracers \cite{Harris:1965} and a large number of studies have then been 
devoted to that problem.\cite{Lizana/Ambjornsson:2008, Taloni/etal:2010, Ryabov/Chvosta:2011, Sanders/Ambjornsson:2012, Lucena/etal:2012, Ryabov:2013, Euan-Diaz/etal:2015, Leibovich/Barkai:2013, Ooshida/etal:2018, Wittmann/etal:2021} 

Recently it has been shown that the single-file constraint gives rise also to intriguing collective transport properties of Brownian particles in periodic potentials.\cite{Lips/etal:2018, Lips/etal:2019, Lips/etal:2020} 
For hardcore interacting particles, this model was termed the Brownian asymmetric simple exclusion process (BASEP), as it could be considered as a 
generalization of the asymmetric simple exclusion process 
(ASEP),\cite{Derrida:1998, Schuetz:2001, Chou/etal:2011}  which is a paradigmatic lattice model for nonequilibrium dynamics.
In the BASEP, the transport properties depend sensitively
on the particle diameter $\sigma$. A rich variety of current-density relations was found, which are known also as
``fundamental diagrams''.\cite{Schadschneider/etal:2010}
This rich variety causes complex diagrams of many nonequilibrium phases to emerge that are characterized 
by different types of steady states
appearing in open systems coupled to particle reservoirs. 

\begin{figure*}[t!]
\includegraphics [scale = 1]{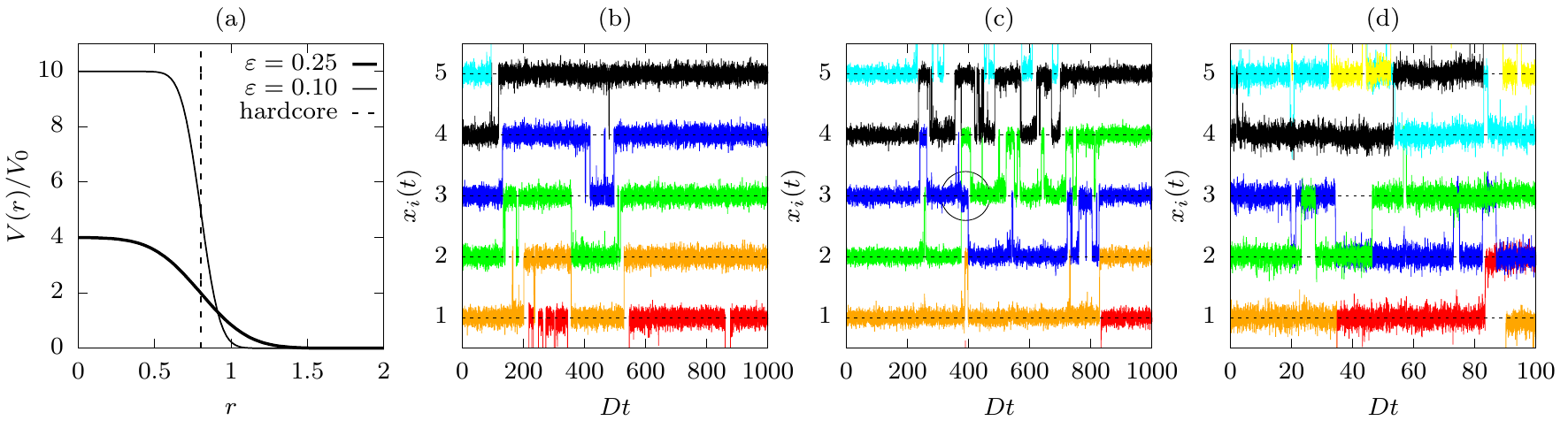}
 \caption{(a) Interaction potential for soft particles of size $\sigma=0.8$ for two values of the softness parameter $\varepsilon$. The dashed line represents the hardcore potential in the limit $\varepsilon\to0$.
 (b)-(d) Trajectories of the driven Brownian motion through the cosine potential in Eq.~\eqref{eq:potential} with the barrier height $U_0=6k_{\rm B}T$
 for a mean particle density $\bar\rho=0.8$ and the particle size $\sigma=0.8$ under a constant drag force $f/k_{\rm B}T=0.2$. 
 The amplitude $V_0$ of the interaction potential is $V_0=k_{\rm B}T$.
 Panel (b) is for hard particles (BASEP), (c) is for soft particles with $\varepsilon=0.1$ (low passing rate), and (d) is for soft particles with 
 $\varepsilon=0.25$ (high passing rate). The circle in panel (c) marks a crossing of two particles.}
\label{fig:potential-trajectories}
\end{figure*}

Here, we study how collective transport properties are influenced 
if the interaction potential is not the ideal hardcore one but 
exhibits some softness.  This applies, for example,
to the motion of interpenetrating macromolecules such as polymer coils in a narrow channel, or one can
imagine hard-sphere colloidal particles to move in a channel slightly larger than their size.
A further example is that of interacting colloids, where the solvent is not perfectly tuned
to yield a hardcore interaction but leads to a soft shell at the colloidal's surface.\cite{Lewis:2000}
Instead of considering such situations in detail, we are interested here in a more generic treatment, where we focus on a rectangular barrier  interaction potential with smoothed barrier steps.

Specifically, the particles interact via the pair potential
\begin{equation}
V(r)=\frac{V_0}{\varepsilon[1+\mathrm{erf}(\sigma/\sqrt{2}\lambda\varepsilon)]}\, \mathrm{erfc}\left(\frac{r-\sigma}{\sqrt{2}\lambda\varepsilon}\right)\,,
\label{eq:interaction-potential}
\end{equation} 
where erf($\cdot$) and erfc($\cdot$) are the error and the complementary error functions defined as\cite{Abramowitz/Stegun:1972} 
${\rm erf}(z)= \left(2/\sqrt{\pi} \right) \int_0^z \dd t\, e^{-t^2}$, and ${\rm erfc}(z)=1-{\rm erf}(z)$.
The potential is shown in Fig.~\ref{fig:potential-trajectories}(a). The variable $r$ is the distance between the center positions of two particles, and $\sigma$ characterizes the particle size.
The wavelength $\lambda$ of the periodic potential appears in Eq.~\eqref{eq:interaction-potential} just as a length scale. In fact, we will use
it as the length unit in the following, i.e.\ $\lambda=1$.
The parameter $\varepsilon>0$ is dimensionless and allows us to control the softness of the interaction. 
We refer to it as the softness parameter. 
With decreasing $\varepsilon$, the barrier edges become sharper  and the interaction strength $V_0/\varepsilon$ increases; see
Fig.~\ref{fig:potential-trajectories}(a).
In the limit $\varepsilon\to0$, the hard-sphere potential for particles with diameter $\sigma$ is recovered. 
We will speak about ``hard particles''  for $\varepsilon=0$ and about ``soft particles'' if $\varepsilon>0$.

To describe the collective particle transport, we focus on current-density relations,
which we determine by extensive Brownian dynamics simulations.
To explain our findings, we introduce an effective particle size for soft-particles based 
on equilibrium properties.
This method bears some resemblance to an approach in the theory of simple fluids in equilibrium. 
Our effective size method applies to nonequilibrium transport properties. It relies on the idea to
use the system of hardcore interacting particles as a reference system for predicting collective dynamics of soft particles.

\section{Model}

We consider an external cosine potential
\begin{equation}
U(x) = \frac{U_0}{2} \cos \left(\frac{2 \pi x}{\lambda} \right)
\label{eq:potential}
\end{equation}
with wavelength $\lambda=1$ and potential barriers $U_0$ between wells 
much larger than the thermal energy $k_{\rm B}T$. We use a fixed ratio $U_0/k_{\rm B}T=6$. 
The amplitude $V_0$ of the interaction potential in Eq.~\eqref{eq:interaction-potential} is set to $k_{\rm B}T$.
A constant drag force $f$ is acting on all particles with $f\lambda/k_{\rm B}T=0.2$.
The length $L$ of the system is equal to $100$, and periodic boundary conditions are applied.
In the steady state, the particles are thus moving along a ring with a mean velocity in the direction of $f$.

For the pair potential in Eq.~\eqref{eq:interaction-potential},
the interaction force of a particle $j$ at position $x_j$ exerted on a particle $i$ at position $x_i$ is
\begin{align}
&f^{(2)}(x_i,x_j)=-\frac{\partial V(|x_i-x_j|)}{\partial x_i}\nonumber\\
&=\frac{\sqrt{2}V_0}{\sqrt{\pi}\varepsilon^2[1+\mathrm{erf}(\sigma/\sqrt{2}\varepsilon)]}\\
&\hspace{7em}{}\times\frac{x_i-x_j}{|x_i-x_j|}\exp(-\frac{(|x_i-x_j|-\sigma)^2}{2\varepsilon^2})\,.\nonumber
\label{eq:f(r)}
\end{align}
In the limit $\varepsilon\to0$ this force approaches a delta-function at the particle distance $\sigma$ with 
amplitude $V_0/\varepsilon$. We will consider two values of $\varepsilon$ in the following, namely $\varepsilon=0.1$
and $\varepsilon=0.25$ corresponding to a low and high passing rate, respectively. The interaction potential for these two $\varepsilon$ values is displayed in Fig.~\ref{fig:potential-trajectories}(a). 

The driven Brownian motion of $N$ particles is described by the Langevin equations 
\begin{equation}
\frac{\dd x_i}{\dd t}=\mu\left(f -U'(x_i) + f^{\rm int}_i\right) + \sqrt{2D}\,\xi_i(t)\,,\hspace{0.5em}i=1,\ldots,N\,,
\label{eq:langevin}
\end{equation}
where $\mu$ is the particle mobility, $D=k_{\rm B}T\mu$ is the diffusion coefficient, and $\xi_i(t)$ are the Gaussian white noise processes with 
zero mean and correlation functions $\langle \xi_i(t) \xi_j(t') \rangle = \delta_{ij}\delta(t - t')$;
$f^{\rm int}_i=\sum_{j\ne i}f^{(2)}(x_i,x_j)$ is the total interaction force acting on particle $i$
and $U'(\cdot)$ is the derivative of the external potential.

\begin{figure*}[t!]
\includegraphics [width=\textwidth]{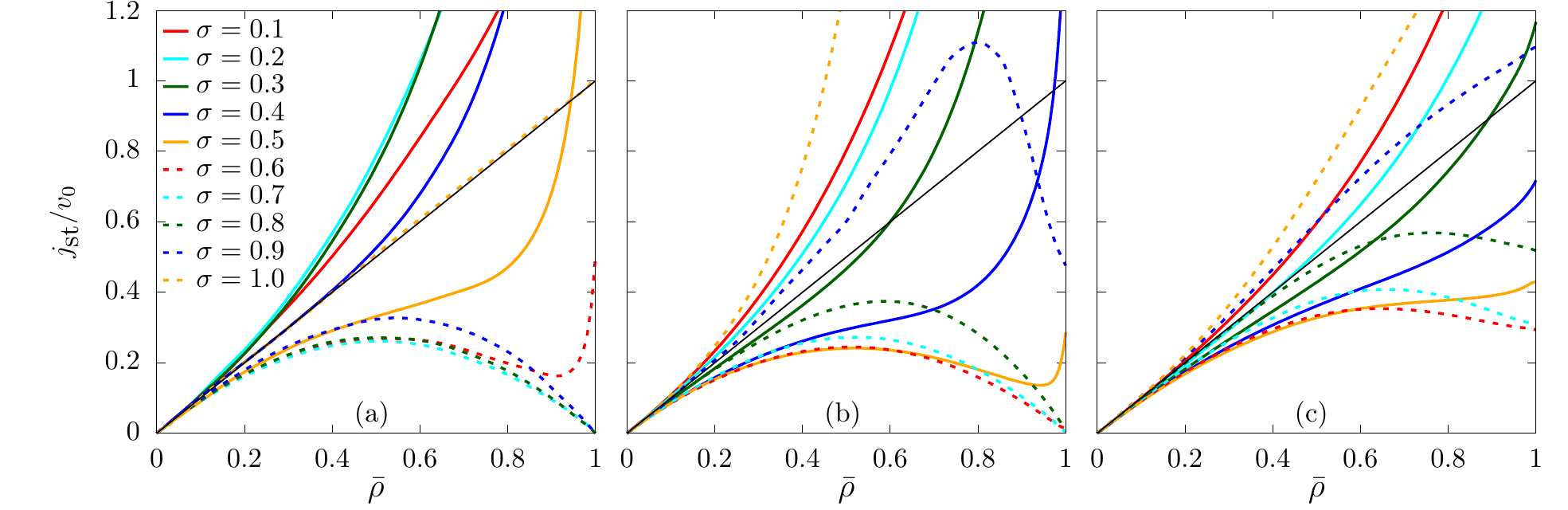}
\caption{Simulated current-density relations for various particle sizes. 
Panel (a) is for hard particles, and panels (b) and (c) are for soft particles with $\varepsilon=0.1$ and $\varepsilon=0.25$, respectively. The legend
in (a) applies to all panels. The current is normalized to the mean velocity $v_0$ of a single particle, and 
the thin solid line with slope 1 represents the behavior of independent particles in all panels.  }
\label{fig:j-rho-BDS}
\end{figure*}

The mean number density of particles is
\begin{equation}
\bar\rho=\frac{N}{L}\,.
\label{eq:barrho}
\end{equation}
Because we defined $\lambda$ as our length unit, we can regard $\bar\rho$ also as the filling factor $N\lambda/L$, i.e.,\
 the mean number of particles per potential well. 

We solved the Langevin equations~\eqref{eq:langevin} by applying the Euler discretization scheme with time steps 
$D\Delta t=10^{-5}-10^{-4}$. It was checked that our results are neither affected by the time step nor by the finite system length.
For treating the hardcore interaction ($\varepsilon=0$), we followed the procedure proposed in Ref.~\onlinecite{Scala:2012} as 
described in detail in Ref.~\onlinecite{Lips/etal:2019}.

Figures~\ref{fig:potential-trajectories}(b)-(d) show representative particle trajectories for hard particles [BASEP, panel (b)], and soft
particles for low and high passing rates [panels (c) and (d)]. In all figures, the density is $\bar\rho=0.8$ and the particles have size $\sigma=0.8$.
Following the trajectories, we see jump-like transitions
between the potential wells, because the thermal energy $k_{\rm B}T$ is much smaller than the potential barrier $U_0$.
While the particles keep their ordering (single-file motion) in Fig.~\ref{fig:potential-trajectories}(b), one particle crossing can be seen in the time window $1000/D$ in Fig.~\ref{fig:potential-trajectories}(c). The corresponding event is marked by a circle. In Fig.~\ref{fig:potential-trajectories}(d),
many more particle crossings occur even in a ten times smaller time window. In addition, we observe many particle overlaps as reflected in overlapping colors. This demonstrates that the potential for $\varepsilon=0.25$ is very soft.

We next discuss current-density relations (fundamental diagrams) in the dependence of the particle size $\sigma$ and mean density $\bar\rho$ for two 
softness parameters $\varepsilon=0.1$ and $\varepsilon=0.25$.

\section{Dependence of currents on density and particle size: Simulations}
\label{sec:simresults}

For independent particles, the current does not depend on $\sigma$ and is given by $j_0(\bar\rho)=v_0\bar\rho$, where
$v_0=v_0(f)$ is the mean velocity of a particle, when it is dragged through the periodic potential $U(x)$ by the force $f$. This mean
velocity is known analytically\cite{Ambegaokar/Halperin:1969} and we use it to normalize
simulated currents.
Normalized currents $j_{\rm st}(\bar\rho,\sigma)/v_0$ are shown in 
Figs.~\ref{fig:j-rho-BDS}(a)-(c) as a function of the density for various fixed particles sizes $\sigma$.
The current-density relations in Fig.~\ref{fig:j-rho-BDS}(a) are for the hard particles ($\varepsilon=0$), and those in 
Figs.~\ref{fig:j-rho-BDS}(b) and \ref{fig:j-rho-BDS}(c) are for the soft particles with $\varepsilon=0.1$ and $\varepsilon=0.25$, 
respectively. The solid line with slope 1 in all graphs marks the current for independent particles.

The current-density relations for the BASEP in Fig.~\ref{fig:j-rho-BDS}(a) have been investigated in our previous studies and serve as a 
reference to discuss the impact of the softness of the potential. In the BASEP, the change in the current with density for varying $\sigma$ 
can be attributed to a barrier reduction and blocking effect that compete with each other. 

The blocking effect is dominant for particle sizes in the range of $0.65\lesssim\sigma\lesssim 0.8$. Transitions of such particles 
into a neighboring potential well are frequently blocked by a particle already residing inside the neighboring well. This leads to the 
current suppression with density (jamming-like behavior) and hence to fundamental diagrams reminiscent of that of the ASEP.

Outside the ASEP-like range, the behavior is more complex and was explained in detail in 
Refs.~\onlinecite{Lips/etal:2018, Lips/etal:2019}. 
The particle current is in particular strongly influenced by the barrier reduction effect. The barrier reduction occurs if more 
than one particle are occupying a single potential well. Such particles are pushing each other away from the potential well minimum 
towards regions of high potential energy. Accordingly, the potential barrier for a transition into the neighboring well is effectively reduced. 
The barrier reduction effect thus leads to a current increase with density stronger than for noninteracting particles. 

In addition, there is a scaling transformation for collective transport quantities under a rescaling of both 
$\bar\rho$ and $\sigma$. For the current, this reads
\begin{equation}
j_{\rm st}(\bar\rho,\sigma)=(1-m\bar\rho)j_{\rm st}\left(\frac{\bar\rho}{1-m\bar\rho},\sigma-m\right)\,,
\label{eq:j-trafo}
\end{equation}
where $m=\mathrm{int}(\sigma)$ is the integer part of $\sigma$. Due to Eq.~\eqref{eq:j-trafo}, the current behavior for
$\sigma\ge1$ can be inferred from that for $\sigma<1$.
The relation also implies the surprising result that the current for 
$\sigma=1$ is equal to that of independent particles; see the respective simulated line in Fig.~\ref{fig:j-rho-BDS}(a). 

Equation~\eqref{eq:j-trafo} tells us that the current behavior drastically changes if $\sigma$ is increased only very little
from $\sigma<1$ to $\sigma>1$. For $\sigma<1$, the current becomes extremely small 
for $\bar\rho\to1$ (the dominant blocking effect) while for $\sigma>1$, the system actually corresponds to
one with a small particle size $(\sigma-1)$, where the barrier reduction effect prevails 
(large currents, including the current in the limit $\bar\rho\to1$).

For soft particles, the scaling transformation does not hold, because in deriving 
Eq.~\eqref{eq:j-trafo}, it is assumed that the particles cannot overlap. 
Nevertheless, the relation \eqref{eq:j-trafo} is very useful for understanding the transport behavior of soft particles.
This will be shown in  Sec.~\ref{subsec:application-theory-low-passing}.

\subsection{Low passing rate}
\label{subsec:j-rho-low-passing}
For discussing how the current-density relations change for the soft particles, let us first focus on the case of low passing rates displayed
in Fig.~\ref{fig:j-rho-BDS}(b) ($\varepsilon=0.1$). The rather complex behavior of $j_{\rm st}(\bar\rho,\sigma)$ in this figure can be described as follows.

For a small $\sigma\lesssim0.3$, in analogy with the behavior in Fig.~\ref{fig:j-rho-BDS}(a), the currents
are dominated by the barrier reduction effect, which leads to an enhancement of $j_{\rm st}(\bar\rho,\sigma)$ 
in comparison to independent particles 
[for $\sigma=0.3$, the current is slightly smaller than $v_0\bar\rho$ in Fig.~\ref{fig:j-rho-BDS}(b) for $\bar\rho\lesssim0.6$]. 
The barrier reduction effect occurs because two (or more) particles can
be in the same potential well. Their average positions inside the well then are displaced from the minimum, leading to a reduced barrier for surmounting the saddle points to the neighboring well. As the fraction of multiple occupied wells increases with $\bar\rho$, the 
enhancement of the current becomes stronger. This is reflected in the upward bending of the current curves
for a small sigma in Fig.~\ref{fig:j-rho-BDS}(b) (and also for $\sigma=0.4$ and 0.5 at large $\bar\rho$).

For intermediate $0.4\lesssim\sigma\lesssim0.8$, the current-density relations in Fig.~\ref{fig:j-rho-BDS}(b)
become influenced and eventually dominated by the blocking effect, which causes the current to become reduced in comparison to
that of independent particles. The blocking effect is well known in lattice models such as the ASEP 
\cite{Derrida:1998, Schuetz:2001, Chou/etal:2011}  and occurs because the motion of a particle from one well to the next
is strongly hindered if the neighboring well is already occupied by a particle. 
In fact, $j_{\rm st}(\bar\rho,\sigma)/v_0$ is close to the parabolic form $\bar\rho(1-\bar\rho)$ of the ASEP for $\sigma=0.6$ and 0.7
in  Fig.~\ref{fig:j-rho-BDS}(b). 

A closer inspection of the current-density relations for both small and intermediate particle sizes in Figs.~\ref{fig:j-rho-BDS}(a) and \ref{fig:j-rho-BDS}(b) reveals an 
interesting common pattern. The relations in Fig.~\ref{fig:j-rho-BDS}(b) appear to be very similar to those in Fig.~\ref{fig:j-rho-BDS}(a) if
$\sigma$ is shifted by about $0.1$.  For example, the line for $\sigma=0.4$ in 
Fig.~\ref{fig:j-rho-BDS}(b) corresponds to the line for $\sigma=0.5$ in Fig.~\ref{fig:j-rho-BDS}(a).
More generally, we can say that the lines for $\sigma=0.1-0.8$ in 
Fig.~\ref{fig:j-rho-BDS}(b) correspond to the lines for $\sigma=0.2-0.9$ in Fig.~\ref{fig:j-rho-BDS}(a).
This suggests that the currents $j_{\rm st}(\bar\rho,\sigma)$ for the soft particles in the respective $\sigma$ regime
can be described by that of the hard particles 
with an effective particle size 
\begin{equation}
\sigma'=\sigma'(\bar\rho,\sigma)\,,
\label{eq:sigmaprime}
\end{equation}
where the dependence of $\sigma'$ on $\bar\rho$ should be weak.

Remarkably, this similarity of the current-density behavior is no longer found for large $\sigma\gtrsim0.9$. The curve for $\sigma=1$ displays an upward bending for the range of densities $\bar\rho\lesssim0.5$ covered in the figure, as if the blocking effect is irrelevant.
One may conjecture that this curve can be interpreted in terms of an effective particle size $\sigma'$ larger than 1, 
which according to Eq.~\eqref{eq:j-trafo}
would correspond to a small particle size. However, as discussed above,  Eq.~\eqref{eq:j-trafo} can no longer be assumed to be valid for soft particles. 

Even more surprising is the curve for $\sigma=0.9$, which seems to have no counterpart in Fig.~\ref{fig:j-rho-BDS}(a). The following question arises: Is it possible to develop a procedure to determine an effective particle size as conjectured in Eq.~\eqref{eq:sigmaprime} by which we can understand the results for soft particles based on that for hard particles? If this is true, the BASEP could 
serve as a reference system for driven Brownian transport through periodic potentials, similar to the 
hard-sphere fluid constituting a proper reference system in the equilibrium theory of simple fluids.\cite{Hansen/McDonald:1991}

\subsection{High passing rate}
\label{subsec:j-rho-high-passing}
The current-density relations for the high passing rate in Fig.~\ref{fig:j-rho-BDS}(c) show features partly analogous to those in Fig.~\ref{fig:j-rho-BDS}(b) for the low passing rate, but with much weaker sensitivity to $\sigma$. The higher probability of passing apparently
causes both the barrier enhancement and the blocking effect to become weaker. As a consequence, the currents deviate 
less from that of independent particles. 
The currents for all $\sigma$ exhibit appreciable values in the limit $\bar\rho\to1$ and fundamental diagrams reminiscent of that of the
ASEP do no longer appear. The weakening of the blocking is responsible also for the fact that the curve for $\sigma=0.9$ does no longer show a maximum followed by a strong decrease with density as in Fig.~\ref{fig:j-rho-BDS}(b).

\section{Theoretical approaches}
\label{sec:theoretical-approaches}

\subsection{Effective size method}
This approach rests on the idea that the BASEP can serve as a reference system: the current of the soft particles
is given by that of the BASEP with the same particle density and an effective hardcore diameter $\sigma'$. To define $\sigma'$, we
require the equilibrium density profile for the softcore interacting particles to agree as closely as possible with that
of the corresponding hardcore interacting ones. 

The equilibrium density profile for the hard-sphere interacting particles follows from minimizing 
the density functional\cite{Percus:1976}
\begin{subequations}
\begin{align}
&\Omega_{\rm hc}(\sigma_{\rm hc};[\varrho])\label{eq:percus}\\
&=\int\limits_0^1 \dd x\, \varrho(x) 
\biggl\{U(x)-\mu_{\rms ch} 
-k_{\rm B} T \left[ 1-\ln\left(
\frac{\varrho(x)}{1-\eta(x,\sigma_{\rm hc})} \right) \right]\biggr\}\,,\nonumber
\end{align}
where
\begin{equation}
\eta(x,\sigma_{\rm hc})=\int\limits_{x-\sigma_{\rm hc}}^x\dd y\,\rho(y)\,.
\end{equation}
\end{subequations}
The determining equation for $\sigma'$ is given by
\begin{equation}
\sigma'(\bar\rho,\sigma)=\argmin_{\sigma_{\rm hc}}\left\{\sup_{0\le x\le1}\left|\frac{\delta\Omega_{\rm hc}(\sigma_{\rm hc};[\varrho_{\rm eq}])}{\delta\rho_{\rm eq}(x)}\right|\right\}\,.
\label{eq:sigmaprime-determ}
\end{equation}
This means that we insert the equilibrium density $\rho_{\rm eq}(x)$ of the soft particles 
with size $\sigma$ and mean density $\bar\rho$
into the 
functional derivative of $\Omega_{\rm hc}$
and vary $\sigma_{\rm hc}$ until we find that $\sigma_{\rm hc}=\sigma'$, where 
$|\delta\Omega_{\rm hc}(\sigma_{\rm hc};[\varrho_{\rm eq}])/\delta\rho_{\rm eq}(x)|$
deviates by the smallest amount from zero in the interval $0\le x\le1$.

Having determined $\sigma'(\bar\rho,\sigma)$ in this manner, the current is given by
\begin{equation}
j_{\rm st}(\bar\rho,\sigma)=j_{\rm st}^{\rms BASEP}(\bar\rho,\sigma')\,, 
\end{equation}
where $j_{\rm st}^{\rms BASEP}$ is the stationary current for the BASEP. 

\subsection{Approximation of zero mean interaction force  (AZMIF)}
The continuity equation for the local density $\rho(x,t)=\langle \sum_{i=1}^N \delta(x-x_i(t)) \rangle$ reads \cite{Lips/etal:2019}
\begin{align}
\frac{\partial \rho(x, t)}{\partial t} &= -\frac{\partial j(x,t)}{\partial x} \label{eq:continuity}\\ 
&\hspace*{-3em}= -\frac{\partial}{\partial x}\left[\mu\left(f-U'(x) + f^{\rm int}(x,t)\right)\rho(x,t) - D\frac{\partial\rho(x,t)}{\partial x}\right]\,,
\nonumber
\end{align}
where 
\begin{equation}
f^{\rm int}(x,t) 
= \frac{1}{\varrho(x,t)} \int_0^L \dd y\, f^{(2)}(x,y)\varrho^{(2)}(x, y, t) 
\label{eq:mean-interaction-force}
\end{equation}
is the local mean interaction force and
$\varrho^{(2)}(x, y, t)=\langle \sum_{i=1}^N \sum_{k=1, k \neq i}^N
\delta(x-x_i(t))\delta(y-x_k(t)) \rangle$ is the two-particle density.

In the steady-state, the density profile and the mean interaction force are time-independent, $\rho(x,t)=\rho_{\rm st}(x)$ and
$f^{\rm int}(x,t)=f_{\rm st}^{\rm int}(x)$. The
current becomes both time-independent and homogeneous, $j(x,t)=j_{\rm st}(\bar \rho, \sigma)$.
From Eq.~\eqref{eq:continuity} one then derives\cite{Lips/etal:2019}
\begin{subequations}
\label{eq:jst-exact}
\begin{equation}
j_{\rm st}(\bar \rho, \sigma) = \frac{\displaystyle f + \overline{f_{\rm st}^{\rm int}}}{\displaystyle\int_0^1\frac{\dd x}{\rho_{\rm st}(x)}}\,,
\label{eq:jst-exact-a}
\end{equation}
where
\begin{equation}
\overline{f_{\rm st}^{\rm int}}=\int_0^1\!\!\dd x\, f_{\rm st}^{\rm int}(x)
\label{eq:jst-exact-b}
\end{equation}
\end{subequations}
is the period-averaged mean interaction force.

In the linear response limit for small  drag force $f$, Eq.~\eqref{eq:jst-exact-a} becomes
\begin{subequations}
\label{eq:j-linresp}
\begin{align}
j_{\rm st}(\bar \rho, \sigma) &= \frac{1 + \alpha}{\displaystyle\int_0^1 \frac{\dd x}{\rho_{\rm eq}(x)}}f\,,
\label{eq:j-linresp-a}\\[0.5ex]
\alpha &= \frac{\partial \overline{f_{\rm st}^{\rm int}}}{\partial f}\Big|_{f=0}\,,\label{eq:j-linresp-b}
\end{align}
\end{subequations}
where $\rho_{\rm eq}(x)$ is the equilibrium density profile for $f=0$. 

The influence of the mean interaction force on the current in Eq.~\eqref{eq:j-linresp-a} is given by the factor $\alpha$, which depends on $\bar\rho$ and $\sigma$. To determine this factor would require a calculation of the two-particle density $\rho^{(2)}_{\rm st}(x,y)$ in the stationary
state (for small $f$). This is a very challenging task. 

An approximation is obtained by neglecting the impact of $\overline{f_{\rm st}^{\rm int}}$ on $j_{\rm st}(\bar \rho, \sigma)$, corresponding to the 
setting $\alpha=0$ in Eq.~\eqref{eq:j-linresp-a}. We referred to this approach as the ``small-driving approximation'' in our 
former studies,\cite{Lips/etal:2018, Lips/etal:2019, Lips/etal:2020} but use the more precise designation ``approximation of zero mean interaction force''  (AZMIF) here.

\section{Application of theoretical approaches and insights from effective size method}
\label{sec:application-theory}
For the BASEP,\cite{Lips/etal:2019} we have found earlier that the AZMIF could capture the variation of currents with density qualitatively and that
a fair quantitative agreement was obtained for small and large particle sizes ($\sigma\lesssim0.3$ and $\sigma\gtrsim0.7$). 
For $\sigma$ around half the wavelength of the periodic potential, the AZMIF gave a less good prediction. The theoretical description could not be 
improved  by including information on pair correlation functions, when we extended the theoretical treatment by employing the dynamic density 
functional theory (DDFT). We applied the DDFT also to the soft particle system with a scheme as discussed in Ref.~\onlinecite{Lips/etal:2019}. Again, 
the AZMIF and DDFT gave overall similar results and we therefore discuss in the following the outcomes of the AZMIF and effective size method.

The reason for the less good predictions of the AZMIF and DDFT for particle sizes $\sigma\simeq0.5$ is that the absolute value of the
period-averaged
mean interaction force $\overline{f_{\rm st}^{\rm int}}$ is significantly larger for $\sigma\approx 0.5$ than for
small and large particle sizes.\cite{Lips/etal:2019}
For the softcore potential we found that the impact of $\overline{f_{\rm st}^{\rm int}}$ for $\sigma\simeq0.5$
is also significant, implying that the AZMIF and DDFT do not provide a good quantitative descriptions as well. This is reflected by
the deviations of the thin yellow dashed lines (AZMIF) from the simulated data (symbols) in
Figs.~\ref{fig:j-comparison-eps01}(a) and \ref{fig:j-comparison-eps025}(a).

Our findings suggest, however, that for the low passing rate 
it is sufficient to develop a better theory for $\overline{f_{\rm st}^{\rm int}}$ for the BASEP, as the impact of 
$\overline{f_{\rm st}^{\rm int}}$ on the currents for the softcore potential is well captured by the effective size method,
see the comparison (thick dashed lines) with the simulated data (symbols) in Fig.~\ref{fig:j-comparison-eps01}(a).

Let us now discuss the current-density relations and comparisons with their theoretical descriptions for the two cases of low and high passing rates in more detail.

\subsection{Low passing rate}
\label{subsec:application-theory-low-passing}
\begin{figure}[t!]
\includegraphics [width=\columnwidth]{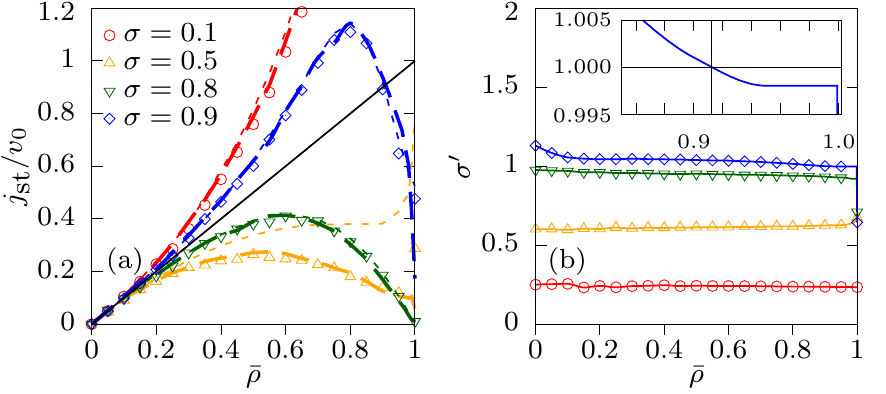}
\caption{(a) Comparison of current-density relations 
predicted by the AZMIF (thin dashed lines) and the effective size method (thick dashed line)
with the simulated data (symbols) for the low passing rate ($\varepsilon=0.1$) and four representative particle sizes.
The solid line with slope 1 indicates the behavior for independent particles. (b) Effective particle sizes calculated from
Eq.~\eqref{eq:sigmaprime-determ} for the data shown in (a)  [same symbols and line colors; see legend in part (a)]. The inset shows a zoomed-in view of the 
curve for $\sigma=0.9$ and $\bar\rho$ close to 1.}
\label{fig:j-comparison-eps01}
\end{figure}

For the case of low passing ($\varepsilon=0.1$), the behavior of $j_{\rm st}(\bar\rho,\sigma)$ predicted by the AZMIF (thin dashed lines) and effective size method (thick dashed lines) is shown in Fig.~\ref{fig:j-comparison-eps01}(a) in comparison with the simulated data.
Similarly as in the BASEP, the AZMIF provides a good description for small and large particle sizes, 
but around $\sigma=0.5$ there appear pronounced differences, see the yellow symbols
and corresponding thin dashed line in  Fig.~\ref{fig:j-comparison-eps01}(a). In contrast, the effective size method gives a good description 
for all particle sizes.

The effective sizes $\sigma'=\sigma'(\bar\rho,\sigma)$ are shown in Fig.~\ref{fig:j-comparison-eps01}(b) as a 
function of $\bar\rho$ for the same $\sigma$ as in Fig.~\ref{fig:j-comparison-eps01}(a).
Note that the $\sigma'$ were determined from Eq.~\eqref{eq:sigmaprime-determ}
by using only equilibrium properties of the system. For 
$\sigma\lesssim0.8$, as we expected in Sec.~\ref{subsec:j-rho-low-passing},
the effective size $\sigma'$ is by about $0.1$ larger than $\sigma$. The variation with 
$\bar\rho$ is weak, except in the limit $\bar\rho\to1$ at large $\sigma$.

The effective size method allows us now to understand the current-density relation for $\sigma=0.9$ in 
Fig.~\ref{fig:j-rho-BDS}(b), which has no counterpart in the current-density relations of the BASEP for any fixed particle size.
As we discussed in Sec.~\ref{sec:simresults}, for $\sigma'$ close to 1, even weak variations in $\sigma'$ with $\bar\rho$
can have a huge impact on the current-density relation. The inset in Fig.~\ref{fig:j-comparison-eps01}(b) shows a decrease of
$\sigma'$ from $\sigma'>1$ to $\sigma'<1$, where $\sigma'=1$ at $\bar\rho_\times\simeq0.91$. This means that
the current of the soft particles should be dominated by the barrier-reduction effect for $\bar\rho<\bar\rho_\times$, 
equal to that of independent particles for $\bar\rho\simeq\bar\rho_\times$, and governed by the blocking effect for $\bar\rho>\bar\rho_\times$. Indeed,
the fundamental diagram for $\sigma=0.9$ in Fig.~\ref{fig:j-rho-BDS}(b) [or Fig.~\ref{fig:j-comparison-eps01}(a)] 
shows the corresponding features. 
In the limit $\bar\rho\to1$, the current for the soft particles is not approaching zero (or very small values) as in the BASEP [see Fig.~\ref{fig:j-rho-BDS}(a)] because the blocking effect is weaker. Surprisingly, this different limiting behavior for $\bar\rho\to1$ is captured qualitatively by a jump-like decrease of $\sigma'$ in the effective size method; see the inset in Fig.~\ref{fig:j-comparison-eps01}(b).

However, contrary to the currents for $\bar\rho<1$, it is very difficult to obtain reliable quantitative predictions of the currents for $\bar\rho=1$ if
$\sigma'\gtrsim0.4$. The reason for this is that the current in the BASEP for $\bar\rho=1$ is strongly decreasing with the particle size  by several orders of 
magnitude for $\sigma\gtrsim0.4$. Hence, even small numerical errors in the determination of
$\sigma'$ can lead to a large change in the predicted current. In addition, the simulated BASEP currents in the respective regime 
are difficult to determine with the required accuracy.

The large soft-particle current for $\sigma=1$ in Fig.~\ref{fig:j-rho-BDS}(b) with its upward bending for $\bar\rho\lesssim0.5$
can be explained by the effective size method as well. In this case, $\sigma'$ is slightly larger than 1 
[not shown in Fig.~\ref{fig:j-comparison-eps01}(b)], implying that the currents
correspond to ones for hard particles with small size $(\sigma'-1)$ according to the transformation \eqref{eq:j-trafo}. Hence, 
the barrier-reduction prevails and we see the corresponding behavior in Fig.~\ref{fig:j-rho-BDS}(b). 

One can explain the dominance of the barrier reduction effect for $\sigma'\gtrsim1$ also without using the 
transformation \eqref{eq:j-trafo}. Imagine two particles with $\sigma'\gtrsim1$ occupying neighboring wells. These cannot be at the minima of the 
potential at the same time, i.e.\ they are on average displaced from the bottom of the wells. Accordingly, they need less energy to overcome the 
saddle point to a vacant neighboring well. Because the number of particles occupying neighboring wells increases with the density, the corresponding 
enhancement of the current becomes stronger with $\bar\rho$, leading to the upward bending of the current-density relations.

\subsection{High passing rate}
\label{subsec:application-theory-high-passing}
\begin{figure}[t!]
\includegraphics [width=\columnwidth]{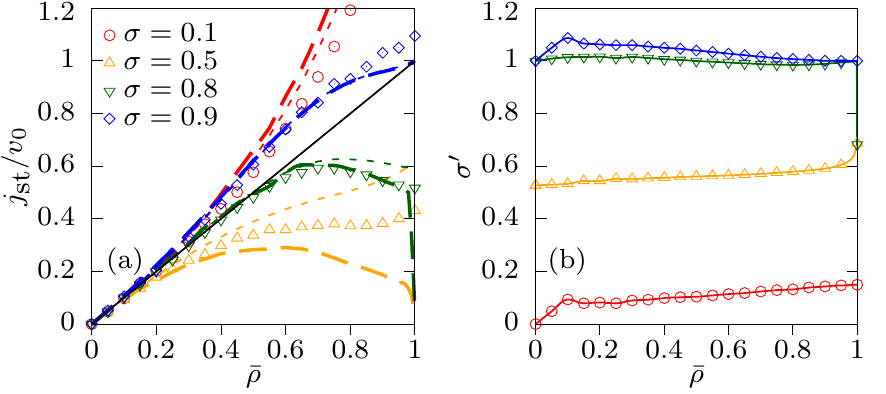}
\caption{(a) Comparison of current-density relations 
predicted by the AZMIF (thin dashed lines) and the effective size method (thick dashed line)
with the simulated data (symbols) for the high passing rate ($\varepsilon=0.25$) and the same particle sizes as in Fig.~\ref{fig:j-comparison-eps01}.
The solid line with slope 1 indicates the behavior for independent particles. (b) Effective particle sizes calculated from
Eq.~\eqref{eq:sigmaprime-determ} for the data shown in (a)  [same symbols and line colors; see the legend in part (a)]. }
\label{fig:j-comparison-eps025}
\end{figure}

For the very soft particles ($\varepsilon=0.25$), one should expect the effective size method to be less appropriate, because the 
method describes modified effective particle sizes. As the method uses the BASEP as a reference system, we expect it to be less 
appropriate if the likelihood of particles overtaking each other becomes relevant. High passing rates will occur in particular
at high densities; see also Fig.~\ref{fig:potential-trajectories}(d). Nevertheless, as shown in Fig.~\ref{fig:j-comparison-eps025},
the effective size method (thick dashed lines) still predicts many features correctly. 
For $\sigma=0.8$, the current-density relation is surprisingly well predicted when $\bar\rho<1$. 
If $\bar\rho=1$, the effective size method suffers from
the same problems as already discussed in Sec~\ref{subsec:application-theory-low-passing}. 

In addition, the 
simulated current-density relations (symbols) for $\sigma=0.1$ and $\sigma=0.9$ are quite well described for $\bar\rho\lesssim0.7$ by the effective size method, whereas at larger $\bar\rho$  stronger deviations are seen. 
For $\sigma=0.5$, significant deviations set in already at $\bar\rho\simeq0.5$. 

These findings can all be reasoned by noting that the effective size method becomes less applicable if
the passing rate of particles is large and if the passing significantly affects the current. 
Passing of soft particles is in particular facilitated by double (or multiple) occupied potential wells, which implies 
that particles with smaller $\sigma$ have higher passing rates at a given density than those with larger $\sigma$.
A higher passing rate, however, is relevant only, when the blocking effect in the corresponding hardcore system
is dominant rather than the barrier reduction effect. In that case, 
the blocking effect is overestimated and the current in the soft particle system 
less strongly reduced with the increasing density than expected from the BASEP reference system.

We thus can conclude the following:
for $\sigma=0.1$, $\sigma'$ is also small [see Fig.~\ref{fig:j-comparison-eps025}(b)], and accordingly, the barrier reduction effect 
dominates the current behavior in the corresponding BASEP. The high passing rate of the soft particles for $\sigma=0.1$ thus is irrelevant 
and the effective size method gives a good description of the current-density relation. For large $\sigma=0.8$ or 0.9, 
the probability of double occupancies of wells becomes significant only at high densities, and therefore the passing rate is 
comparatively low for smaller densities $\bar\rho$, leading to the quite good agreement of the predicted currents 
with the simulated ones. For the intermediate value $\sigma=0.5$, the passing rate of the soft particles is quite large and the effective 
$\sigma'$ is in a regime, where the blocking effect has a comparable or even stronger impact on the current in the BASEP than the barrier reduction effect. As a consequence, quite pronounced deviations are seen between the predicted and simulated currents already at moderate densities $\bar\rho$.

\section{Summary and Conclusions}
\label{conclusion}

We have studied the Brownian motion of particles through 
a cosine potential under a constant drag force as a model for the driven diffusive transport through pore-like structures.
Our aim was to explore how a softness in the particle interaction affects the collective transport properties. To allow for a systematic comparison with hardcore interacting particles, we considered the interaction to be given by a smoothed rectangular barrier potential 
whose softness could be tuned by a single parameter. 

The softness of this potential brings about two features:
a smoothed barrier step and a finite height of the barrier. The smoothed step 
causes particles to partially penetrate each other (penetration effect), and the finite barrier allows particles to pass each other (passing effect).
We focused our analysis on two cases of low and high passing rates. In the case of the low passing rate,
only the penetration effect is essentially present. 

Even if the passing rate is negligible, we found peculiar current-density relations having no counterpart in a hardcore interacting system.
To explain this, we introduced an effective particle size method. In this method, we map a system of softcore interacting particles onto that of 
hardcore interacting ones with the same density and an effective hardcore diameter that is determined based on the equilibrium 
density functional of hard spheres. This effective hardcore diameter depends on both the size and density of the soft particles, and
due to this dependence, the peculiar current-density relations could be well described. We can conclude therefore that the 
effective size method accounts correctly for the impact of the penetration effect on the transport behavior. In addition,
it allows one to interpret the dynamics based on the knowledge about the hardcore interacting reference system. 

If the potential barrier is low, the passing rate of particles can become important. In this case there exist two regimes, one at low density, where the penetration effect prevails, and another one at high density, where the passing effect is dominant. In the low-density regime, the current-density relation can thus be successfully described by our effective size method. In the high-density regime, the effective size method does no longer provide a good quantitative description. Apparently, a different or modified theoretical approach is needed to cope with the collective dynamics in that regime.

The effective size method has been applied here to describe current-density relations as a fundamental aspect of nonequilibrium dynamics. We expect it to apply also to other nonequilibrium properties. Moreover, the method could be a valuable approach for treating soft particles with long-range attractive interactions.\cite{Likos:2001} To determine the effective size in that case, advancements of equilibrium density functional theory for corresponding hardcore interacting systems are useful.\cite{Lutsko:2010, Wittmann/etal:2016} Likewise, mixtures of soft particles may be treated based on density functionals for hard-sphere mixtures.\cite{Vanderlick/etal:1989, Bakhti/etal:2012} As for experimental verification of the findings reported here, it is important to point out that the striking effects in the collective transport can not only be identified in currents but also in the local kinetics of tagged particles.\cite{Ryabov/etal:2019, Vorac/etal:2020}

\begin{acknowledgements}\vspace{-2ex}
Financial support from the Czech Science Foundation (Project No.\ 20-24748J) and the Deutsche Forschungsgemeinschaft (Project No.\ 432123484) is 
gratefully acknowledged. 
\end{acknowledgements} 

\vspace{-3ex}
\section*{Data Availability}\vspace{-2ex}
The data that support the findings of this study are available from the corresponding author upon reasonable request.



%

\end{document}